\documentstyle[aps,preprint]{revtex}

\newcommand{\beq}{\begin{equation}}
\newcommand{\eeq}{\end{equation}}
\newcommand{\bea}{\begin{eqnarray}}
\newcommand{\eea}{\end{eqnarray}}
\def\vev#1{\left\langle #1\right\rangle}
\newcommand{\ra}{\rightarrow}

\begin{document}
\preprint{SNUTP 94-02}
\title{Small Flavor Conserving CP Violation in Superstring Models\footnote{
To appear in the Proceeding of the Third
KEK Topical Conference on CP Violation,
its Implication to Particle Physics and Cosmology} }
\author{Kiwoon Choi}
\address{Department of Physics,
        Chonbug National University\\
        Chonju  560-756, KOREA}

\maketitle

\begin{abstract}
It is well known that supersymmetric models allow
new sources for CP violation that arise from soft
supersymmetry breaking terms. If unsuppressed,
these new CP-violating phases would give too large a
neutron electric  dipole moment.
We discuss a mechanism for suppressing SUSY phases
in string-inspired supergravity models
in which  supersymmetry is assumed to be broken
by the auxiliary components of the dilaton and moduli superfields.
\end{abstract}

\section{Introduction}

Since the first experimental observation of CP violation, the flavor
(strangeness) changing decays of $K$ mesons remain to be
the only place in which the violation has been seen
experimentally. Although not observed yet, flavor conserving
CP violation, e.g. particle electric dipole moments,
is also of great interest since it can provide useful information
on physics beyond the  standard model.

As is well known, the standard model contains only two sources
for CP violation, the Kobayashi-Maskawa (KM) phase $\delta_{\rm KM}$
and the QCD vacuum angle $\bar{\theta}$.
To explain the observed CP violation
in $K$-decays, $\delta_{\rm KM}$ is required to be of order unity.
Even for $\delta_{\rm KM}$ of order unity,  the resulting flavor
conserving CP violation is  negligibly small  because
it is second order in weak interactions
while its CP conserving counterpart  occurs at zeroth order
without resorting to weak interactions.
Note that although $\triangle S=2$ CP-odd $K$-$\bar{K}$
mixing is second order in weak interactions, its CP-even
counterpart is second order also, allowing the experimental
observation of $\epsilon\sim {\rm odd}/{\rm even}\sim 10^{-3}$
where the small value of $\epsilon$ is due to small mixing angles,
{\it not} due to a small $\delta_{\rm KM}$.

Contrary to this, to be compatible with
the current experimental limit
on the neutron electric dipole moment $d_n\leq 10^{-25} e\cdot cm$,
$\bar{\theta}$ is required to be {\it less} than $10^{-9}$ $\cite{strongcp}$.
For $\bar{\theta}=\theta_{\rm QCD}+\theta_{\rm QFD}$,
both $\theta_{\rm QFD}={\rm arg}({\rm det} M_q)$ and
$\delta_{\rm KM}$ originate
from complex Yukawa couplings, and thus
are expected to have the same order of magnitudes.
Then the phenomenological limit on $\bar{\theta}$
requires fine tuning  $\theta_{\rm QCD}$ to cancel $\theta_{\rm QFD}$.
The strong CP problem of this fine tuning
has motivated a variety of extensions
to the standard model which accommodate
a mechanism for setting $\bar{\theta}$ to zero
$\cite{strongcp}$.

In recent years, supersymmetry (SUSY) has emerged as
a leading candidate for physics beyond the standard model
around the weak scale. This is largely due to the fact that
SUSY provides a perturbative
solution to the problem of the quadratic divergence in scalar masses.
Furthermore if SUSY breaking is triggered by nonperturbative
effects, supersymmetry
may also provide an explanation for the large hierarchy
between the weak scale and the Planck scale $\cite{witten1}$.

Although quite attractive in view of the hierarchy problem,
conventional supersymmetric models suffer from several
naturalness problem not shared by the standard model.
For instance, they require a high level of degeneracy in squark and slepton
masses to avoid too large flavor changing neutral currents.
This is hard to be  understood in view of a rather large hierarchy
in quark and lepton masses. Besides this SUSY flavor problem,
there is another naturalness problem,
the ``SUSY phase problem" $\cite{garisto}$, which is the subject
of this talk.
As is well known, supersymmetric models
allow new sources for CP violation
arising from soft SUSY breaking terms.
For superpartner masses around the weak scale,
these new SUSY phases are required to be {\it less}
than $10^{-2}-10^{-3}$ in order to avoid  too large
a neutron electric dipole moment $\cite{garisto}$.
Although not as severe as $\bar{\theta}$,
it is still hard to understand how
theories can give rise to small SUSY phases while giving
the KM phase of order unity.

Clearly the SUSY phase problem
is a problem of SUSY breaking since the relevant phases originate from
soft SUSY breaking terms.
Presently the most popular way to break SUSY
is  to introduce a hidden sector into
the underlying $N=1$ supergravity model $\cite{nilles1}$.
Four dimensional $N=1$ supergravity theories are highly
nonrenormalizable and thus are considered to be an
effective theory of a more fundamental theory which is presumed
to consistently unify the gravity with  particle physics interactions.
Presently string theory is the only known candidate
along this direction $\cite{witten}$.
In this regard, it would be quite interesting to explore
the possibility of small SUSY phases
in string-inspired supergravity models.

Here we wish to discuss a mechanism for
suppressing SUSY phases which would be realized in
string-inspired models if some {\it plausible} conditions
are met $\cite{choi2}$.  The  essential ingredients of the
mechanism are: (i) approximate Peccei Quinn symmetries
nonlinearly realized for the pseudoscalar components of the moduli
superfields which contribute to SUSY breaking by having nonzero
auxiliary components,
(ii) dynamical relaxation of the relative phases
in the nonperturbative superpotential of the dilaton and
moduli superfields.
The conditions for the mechanism to work are not so restrictive,
and are  satisfied in fact  by many
SUSY breaking scenarios in string
theories $\cite{louis1,casas}$.

\section{The small SUSY phase problem}

To begin with,
let us  consider
a generic renormalizable supersymmetric
model  with the effective superpotential
\begin{equation}
W_{\rm eff}=\lambda_{ijk} \Phi_i\Phi_j\Phi_k+\mu_{ij}\Phi_i\Phi_j,
\end{equation}
and the soft breaking terms
\bea
{\cal L}_{\rm soft}&=&
\frac{1}{2}m^2_{ij}\varphi_i\varphi_j^{\ast}+
A_{ijk}\varphi_i\varphi_j\varphi_k \nonumber \\
&&+B_{ij}\varphi_i\varphi_j+\frac{1}{2}m_a\lambda_a\lambda_a
+{\rm h.c.},
\eea
where $\Phi_i$ denote generic  chiral  superfields,
$\varphi_i$ are their scalar components, and  $\lambda_a$ are
the gaugino for the $a$-th
gauge group.

 All the parameters that appear in $W_{\rm eff}$ and
 ${\cal L}_{\rm soft}$ are complex in  general, and thus can give rise to
CP-violating phenomena in principle.
However  physical CP violation occurs only through
CP-odd parameters  which can {\it not}
be eliminated by  reparametrizing field variables.
The hermitian
mass matrix  $m^2_{ij}$ can always be made
to be real and  diagonal by
a unitary transformation of superfields:
$\Phi_i\ra U_{ij}\Phi_j$. One can still perform
an arbitrary phase rotation of superfields,  $U(1)_i:
\Phi_i\ra e^{i\alpha_i}\Phi_i$,
to eliminate some of the phases while
keeping $m^2_{ij}$  real and diagonal.
Clearly the following three classes of phases
\bea
\phi_A&=&\{{\rm arg}(A_{ijk}/\lambda_{ijk})\}, \nonumber \\
\phi_B&=&\{{\rm arg}(B_{ij}/\mu_{ij})\}, \nonumber \\
\phi_C&=&\{{\rm arg}(m_a)\}
\eea
are invariant under $U(1)_i$, and thus
are the candidates for the sources for physical CP violation.
There is a further transformation which allows one of the above
phases to be rotated away.  It is an $R$-transformation
$U(1)_R$:
\beq
\Theta\ra e^{i\alpha}\Theta, \quad \Phi_i\ra \Phi_i,
\eeq
where $\Theta$ denotes the Grassmann coordinate of superspace.
Obviously $\phi_{A,B,C}\ra
\phi_{A,B,C}+2\alpha$ under $U(1)_R$, implying that one of the phases can be
eliminated
by an appropriate $U(1)_R$ transformation.

The above discussion shows that a number of
CP violating phases can arise from soft supersymmetry breaking terms.
Inspired by simple hidden sector supergravity models,
the following universality conditions are often
assumed for the coefficients of  soft terms
renormalized at the Planck scale:
\beq
A_{ijk}=A\lambda_{ijk}, \quad B_{ij}=B\mu_{ij}, \quad m_a=\tilde{m}.
\eeq
In this simple case,  soft supersymmetry breaking terms provide
only {\it two} new reparametrization-invariant CP-violating phases
which can be chosen to be $\cite{hall}$
\bea
\phi&=&\{{\rm arg}(\tilde{m}A^{\ast}), \,
 {\rm arg}(\tilde{m}
B^{\ast})\}\nonumber \\
&=&\{\phi_C-\phi_A, \, \phi_C-\phi_B\}.
\eea

Supersymmetric lagrangian contains a squark-quark-gluino coupling
$I_{q\tilde{q}\lambda}\sim
\tilde{q}^{\ast}T^a q \lambda^a$  which is manifestly
CP-conserving when  written in terms of {\it weak eigenstate}
fields.
Due to soft SUSY breaking terms, the unitary matrix relating
the weak eigenstate quarks to  the mass eigenstate quarks is
different from the unitary matrix  for squarks.
As a result,  the coupling $I_{q\tilde{q}\lambda}$ written
in terms of  {\it mass eigenstate} fields would contain
flavor-conserving but CP-violating piece which depends upon the phase $\phi$.
Radiative effects of this coupling
then induce flavor-conserving  but CP-violating effective operators
of quarks and gluons.
In regard to
the neutron electric dipole moment $d_n$, the operators of interest
are the electric dipole moments (EDM) of light quarks and the chromo-electric
dipole moments (CEDM) of light quarks and gluon:
\bea
{\cal O}_q^E&=&
d^E_q\frac{m_q}{M_W^2}eF_{\mu\nu}\bar{q}i\sigma^{\mu\nu}\gamma_5 q,\nonumber \\
{\cal O}_q^C&=&
d^C_q\frac{m_q}{M_W^2}g_cG^a_{\mu\nu}\bar{q}i\sigma^{\mu\nu}\gamma_5
T^aq,\nonumber \\
{\cal O}_g^C&=&
d^C_g\frac{1}{M_W^2}g_c^3f_{abc}G^a_{\mu\alpha}G^{b\alpha}_{\nu}
G^c_{\rho\sigma}\epsilon^{\mu\nu\rho\sigma}.
\eea

The EDM and CEDM of $q$ are induced by one-loop graphs
involving  the internal squark $\tilde{q}$ and the gluino
$\lambda^a$ attached to the external $q$ via the coupling
$I_{q\tilde{q}\lambda}$,
while the gluon CEDM
is due to two loop graphs involving the internal $t$,
$\tilde{t}$ and $\lambda^a$.
Note that the EDM and CEDM of $q$ change
the chirality of $q$, and thus
are suppressed by the small quark mass $m_q$.
The EDM of light quarks  in supersymmetric models was computed
a decade  ago $\cite{susyedm}$ while the potential importance of
the gluon CEDM was noted rather recently by Weinberg $\cite{weinberg}$.

Without including perturbative QCD corrections,
one can naively estimate the contributions
to $d_n$ from the above three operators.
Since ${\cal O}^C_g$ is a two-loop result while ${\cal O}^{E,C}_q$
are one-loop results,
one roughly expects
$d^E_q\sim d^C_q\sim 16\pi^2 d^C_g$,
implying
\beq
d_n({\cal O}_q^E)\sim d_n({\cal O}_q^C)
\sim \xi d_n({\cal O}_g^C), \nonumber
\eeq
where $\xi=(m_q/M_n)(4\pi/\alpha_c)$ for the neutron mass $M_n$.
Since $\xi$ is roughly of order unity, one needs a more detailed
analysis including QCD corrections in order to
see which of the operators gives a dominant contribution.
Recently the neutron EDM resulting from
the operators of eq.~(7) in supersymmetric models
was reassessed in ref. $\cite{nano}$. The result indicates that
the contribution from
${\cal O}_q^E$ slightly dominates other contributions, yielding
\beq
d_n({\cal O}_q^E)\simeq
10^{-22}
\eta_q
\left(\frac{100 \, {\rm GeV}}{M_S}\right)^2
\sin \phi \, \, e\cdot cm,
\eeq
where $M_S$ denotes the superpartner masses
which are assumed to have a common value, and
$\eta_q=c_q Q^{\rm em}_q m_q/6 \, {\rm MeV}$ for
the electromagnetic charge $Q^{\rm em}_q$
of $q$ and the matrix element $c_q$ which is defined
by $\vev{n|\bar{q}\sigma_{\mu\nu}q
|n}=c_q \bar{n}\sigma_{\mu\nu}n$.
One then has
\beq
\phi\leq 10^{-3} \eta_q^{-1}
 \left( \frac{M_S}{100 \, {\rm GeV}}\right)^2
\eeq
for the  current experimental limit of $d_n\leq 10^{-25} e\cdot cm$.

For the above constraint to be satisfied,
one needs either $\phi$ to be small enough  or
$M_S$ to be significantly larger than $100$ GeV.
For instance, for $M_S=100$ GeV and
$q=u$ for which $\eta_u\simeq 1$,
one has the limit $\phi \leq 10^{-3}$.
If there exists a significant amount of the {\it sea
strange quarks} in the neutron and thus $c_s m_s$ is significantly
greater than $c_u m_u$, whose possibility has been
argued in the literatures $\cite{manohar}$, one would obtain an even
stronger limit $\cite{he}$.
Clearly the bound
on  $\phi$ can be relaxed by assuming that the superpartners
have masses larger than 100 GeV. For instance, if $M_S$
is about a few TeV, $\phi$ can be of order unity.
Although the option of heavy superpartners is still allowed
$\cite{kizukuri}$,
a phenomenologically more desirable
possibility is to have $\phi\leq 10^{-2}-10^{-3}$
while the superpartner masses are remained to be around a few hundreds GeV
$\cite{garisto}$.
It is then hard to understand how theories
can give rise to small SUSY phases while giving
the KM phase of order unity.
The required smallness of $\phi$
is less severe than that of $\bar{\theta}$, but this
problem of small SUSY phase
persists even in models with a mechanism, e.g. axions,
for setting $\bar{\theta}$ to zero.

\section{Some features of string-inspired supergravity}

Let us briefly discuss some features
of string-inspired supergravity models
which will be relevant for our later discussion
of small SUSY phases.
The model we consider here
has the following  properties which are believed to
be generic properties of  superstring vacua.
First of all, the model contains
a hidden sector which provides nonperturbative
dynamics  for SUSY breaking. This nonperturbative hidden sector
generally has a large gauge
group as well as matter fields
that transform nontrivially under the hidden sector gauge group.

The model contains also the dilaton superfield $S$ and
the overall modulus superfield $T$ $\cite{witten}$.
The dilaton scalar component ${\rm Re}(S)$
couples to the gauge kinetic
terms and thus determines the gauge coupling constant
$g^2=1/{\rm Re}(S)$.
The overall modulus
${\rm Re}(T)$ corresponds to
the radius of  the compactified  internal space in the Planck
length unit. Then $1/{\rm Re}(T)$ would  determine  the
coupling constant of the world sheet sigma model for strings
propagating in the internal space.
In fact, $S$ and $T$ can  be considered as  members
of hidden sector in the sense that
they couple to observable sector by a gravitational
strength. However here
we distinguish them from the other type of hidden sector
providing nonperturbative dynamics for SUSY breaking.

Compactified  superstring predicts moduli fields
other than $T$ in general.
In this section, we ignore
such moduli for the sake of simplicity.
The effects of including
other moduli will be discussed later.
Then {\it the key assumption} of the model  is that local
SUSY is broken by the auxiliary $F$-components of $S$
and/or $T$, which are induced essentially
by the nonperturbative  hidden sector
which couples to $S$ and $T$ by a gravitational strength.

A property of $S$ and $T$ which is crucial
for our later discussion is that in perturbation
theory their pseudoscalar components,
the model-independent axion ${\rm Im}(S)$ and
the internal axion ${\rm Im}(T)$,
decouple at zero four momentum $\cite{witten}$.
In spacetime perturbation theory in which
the background gauge field configuration is topologically trivial,
the vertex operators of these axion-like fields  {\it at zero
four momentum}
are given by:
\bea
V_{{\rm Im}(S)}&=&\int d^2\sigma \epsilon^{\alpha\beta}B_{\mu\nu}
\partial_{\alpha}X^{\mu}\partial_{\beta} X^{\nu},
\nonumber \\
V_{{\rm Im}(T)}&=&\int d^2\sigma \epsilon^{\alpha\beta}B_{mn}
\partial_{\alpha}Y^m\partial_{\beta}Y^n,
\eea
where $X^{\mu}$ and $Y^m$ denote the  string coordinates
of the
flat Minkowski spacetime and the internal space respectively,
$B_{\mu\nu}$ is a constant two-form,
and $B_{mn}$ is the K\"{a}hler form of the internal space.
Clearly
$V_{{\rm Im}(S)}$ is a total divergence and thus vanishes.
However
$V_{{\rm Im}(T)}$ vanishes only  for $Y^m$ which
corresponds to
a topologically trivial mapping from the string world sheet
to the internal space.
This  implies that the constant mode of ${\rm Im}(S)$ decouples
in spacetime perturbation theory ignoring Yang-Mills instanton
fluctuations,
while that of   ${\rm Im}(T)$  decouples in a more restrictive
case ignoring both Yang-Mills and
world sheet instanton fluctuations $\cite{dine}$.

The decoupling of the constant modes of ${\rm Im}(S)$ and
${\rm Im}(T)$ means the invariance
under the nonlinear Peccei Quinn (PQ) symmetries:
\bea
U(1)_S:&& S\rightarrow  S+i\alpha_S,
\nonumber \\
U(1)_T:&& T\rightarrow T+i\alpha_T,
\eea
where $\alpha_{S,T}$ are arbitrary real constants.
In the early stage of anomaly free superstring theories,
it has been speculated that the model-independent axion
${\rm Im}(S)$ and/or the internal axion ${\rm Im}(T)$
may solve the strong CP problem $\cite{witten2}$,
barring the cosmological
difficulty associated with  too large decay constants
$\cite{choi}$.
However to be useful for the  strong CP problem,
explicit breaking of $U(1)_S$ or of $U(1)_T$
must be highly dominated
(by a factor larger than $10^9$) by
the gluon anomaly $G\tilde{G}$.
For $U(1)_S$, this would be achieved in an {\it uninteresting} case
that the color $SU(3)$ is the {\it only} confining
non-abelian gauge group of the model.
However in cases with  non-abelian hidden sector
gauge group,
PQ symmetries are broken either by hidden sector Yang-Mills instantons
or by world sheet instantons,
whose effects are much stronger than that of the gluon
anomaly in general. Since non-abelian hidden sector gauge
group is strongly motivated for SUSY breaking,
the axion-like fields ${\rm Im}(S)$ and ${\rm Im}(T)$ are
considered to be {\it irrelevant} for the strong CP problem.
However as we will see in the next section
they can be very useful for solving
another naturalness problem, the small SUSY phase problem.

We already noted that $U(1)_S$  is  broken
by Yang-Mills instantons including those of the hidden sector
gauge group,
and $U(1)_T$ is by both Yang-Mills and  world sheet
instantons.
Since $1/{\rm Re}(S)$ and $1/{\rm Re}(T)$ correspond to
the gauge coupling constant  and  the sigma model coupling constant
respectively, the effects of Yang-Mills instantons
would be suppressed by
$q_S\equiv e^{-b_S S}$ while those of world sheet instantons
are by $q_T\equiv e^{-b_T T}$ where $b_S$ and $b_T$
are some real constants.
For $S$ normalized as $1/{\rm Re}(S)=g^2$ (at string
tree level), one has  $b_S$ of order $4\pi^2$, leading to
$q_S\ll 10^{-3}$ for phenomenologically desirable
$g^2\simeq 1$.
A common normalization of $T$ is
$T\equiv T+i$ for which $b_T=2\pi$.
In the following, we will {\it assume} that  $q_T
= e^{-2\pi T}\leq 10^{-2}-10^{-3}$ or even less,
so that for the estimate of SUSY phases
up to the accuracy of $10^{-2}-10^{-3}$,
PQ-violating effects can be ignored
if they do {\it not} correspond to leading effects, but
just give   small corrections to perturbative PQ-conserving parts.
Of course to justify this assumption, one needs to
determine
the vacuum value of the modulus ${\rm Re}(T)$
by evaluating the effective potential.
This has been done for some orbifold models, yielding
${\rm Re}(T)\simeq 1.2$ $\cite{casas,font}$ for which
$q_T\leq 10^{-3}$. At any rate,   it is quite
conceivable that $q_T\leq 10^{-2}-10^{-3}$ since
$q_T$ is exponentially
small for a moderately large value of ${\rm Re}(T)$.

As is well known, a four-dimensional $N=1$ supergravity
action is characterized by the K\"{a}hler potential $K$,
the superpotential $W$,  and the gauge kinetic function $f_a$
for the $a$-th gauge group.
Let us start with a generic supergravity model containing
nonperturbative hidden sector which provides a seed for
supersymmetry breaking.
By our assumption, the nonperturbative hidden sector
of the model  does not directly develop
SUSY breaking vacuum  values.
Then the effective theory obtained by integrating out the
nonperturbative hidden sector would still have the structure
of $N=1$ supergravity.
The K\"{a}hler potential and
the superpotential of this  effective supergravity can be
expanded in observable chiral superfields $\Phi_i$ as
\begin{eqnarray}
K&=&\tilde{K}+Z_{ij}\Phi_i\bar{\Phi}_j+(Y_{ij}\Phi_i\Phi_j
+{\rm h.c.})+..., \nonumber \\
W&=&\tilde{W}+\tilde{\mu}_{ij}\Phi_i\Phi_j
+\tilde{\lambda}_{ijk}\Phi_i\Phi_j\Phi_k+...
\end{eqnarray}
Here all coefficients in the expansion
are generic functions
of  $S$ and $T$,
and the ellipses stands for the terms of higher order.
Although the wavefunction factor $Z_{ij}$
can have an off-diagonal element in general,
we assume it is diagonal, viz $Z_{ij}=Z_i\delta_{ij}$,
for the sake of simplicity.
At any rate, off-diagonal elements are required to be small,
roughly smaller than   $10^{-2}Z_i$,
to avoid  too large flavor changing neutral currents.

The terms $\tilde{W}$ and $\tilde{\mu}_{ij}\Phi_i\Phi_j$
in the superpotential vanish in
perturbation theory.
Thus they are purely due to
nonperturbative dynamics of the integrated hidden sector.
Note that a nontrivial nonperturbative superpotential $\tilde{W}$
of $S$ and $T$ is essential for SUSY breaking by
nonzero $F$-components (here the indices $I,J=S,T$):
\begin{equation}
\bar{F}_I=e^{\tilde{K}/2}|\tilde{W}|(\bar{\partial}_I\partial_J\tilde{K})^{-1}
(\partial_J\tilde{K}+\partial_J\ln \tilde{W}).
\end{equation}

In string theory,  CP corresponds to a
discrete  gauge symmetry $\cite{gaugecp}$,
which is an element of higher dimensional Lorentz, general
coordinate transformation, and Yang-Mills groups.
As a result, there is no CP-violating bare parameter in string theory
and CP must be
broken spontaneously.
Spontaneous CP violation
would occur  through the complex vacuum values of
superheavy (Kaluza-Klein or stringy) fields and/or through
the vacuum values of light fields.
If CP violation is {\it entirely} due to light fields,
one can study CP violation using
a CP-invariant  effective lagrangian of light fields.
However if some of superheavy  fields develop CP-violating
values,  after integrating out them,
CP  would appear to be explicitly broken
in the effective lagrangian of light fields.
Here we do not make any assumption on the
nature of CP violation, and thus
allow all complex parameters in the K\"{a}hler potential and
the superpotential
to have the phases of order unity in general.

Now  one can integrate  out $S$ and $T$
to obtain the effective lagrangian of observable fields,
including soft SUSY breaking terms induced by
nonzero values of the auxiliary components $F_{S,T}$.
The resulting effective superpotential $W_{\rm eff}$  (see eq.~(1))
is given by
$\cite{louis,ibanez2}$:
\bea
\lambda_{ijk}&=&e^{-i\xi}e^{\tilde{K}/2}\tilde{\lambda}_{ijk},
\nonumber \\
\mu_{ij}&=&\mu_{1ij}+\mu_{2ij}+\mu_{3ij},
\eea
where $\xi={\rm arg}(\tilde{W})$ and
\bea
\mu_{1ij}&=&\lambda_{Nij}\langle{N}\rangle,\nonumber \\
\mu_{2ij}&=&(m_{3/2}-\bar{F}_I\bar{\partial}_I)Y_{ij},\nonumber \\
\mu_{3ij}&=&e^{-i\xi}e^{\tilde{K}/2}\tilde{\mu}_{ij},
\eea
for the gravitino mass  given by $m_{3/2}=e^{\tilde{K}/2}|\tilde{W}|$.
Here  we consider  three possible sources for the bilinear
terms  in $W_{\rm eff}$.
Note that the $\mu_{1ij}$-piece is obtained by replacing the singlet field $N$
(if it exists) which has the trilinear coupling $\lambda_{Nij} N\Phi_i\Phi_j$
by its vacuum value.

For the soft terms of eq.~(2), one finds $\cite{louis,ibanez2}$
\begin{eqnarray}
A_{ijk}&=&\lambda_{ijk}F_I
\partial_I[\ln(e^{\tilde{K}}\tilde{\lambda}_{ijk}/Z_iZ
_jZ_k)],
\nonumber \\
B_{ij}&=&B_{1ij}+B_{2ij}+B_{3ij}, \nonumber \\
m_a&=&\frac{1}{2}g_a^2F_I\partial_If_a,
\end{eqnarray}
where
\begin{eqnarray}
B_1/\mu_1&=&A_{Nij}/\lambda_{Nij}, \nonumber \\
B_2/\mu_2&=&F_I\partial_I[\ln(e^{\tilde{K}/2}\mu_{2ij}/Z_iZ_j)]-m_{3/2},
\nonumber \\
B_3/\mu_3&=&F_I\partial_I[\ln(e^{\tilde{K}}\tilde{\mu}_{ij}
/Z_iZ_j)]-m_{3/2}. \nonumber
\end{eqnarray}

In fact, soft parameters given above correspond
to those renormalized at a scale around the Planck scale.
Thus it is necessary to include renormalization effects
when one applies the above formulae for low energy phenomenology.
For the SUSY phases $\phi_A={\rm arg}(A_{ijk}/\lambda_{ijk})$,
$\phi_B={\rm arg}(B_{ij}/\mu_{ij})$, and $\phi_C={\rm arg}(m_a)$,
one needs to take into account the renormalization group (RG) mixing
with  other potentially large CP-violating phases, particularly with
the KM phase $\delta_{\rm KM}$ which is of order unity,
that occurs at scales between the Planck scale
and the weak scale.
It turns out that the RG mixing with $\delta_{\rm KM}$
gives a negligible  correction, and thus
$\phi_{A,B,C}$ at the weak scale remain to be small
enough as long as
they are less than $10^{-2}-10^{-3}$ at the Planck scale
$\cite{rgrunning}$.

For the supergravity action invariant
under the PQ symmetries $U(1)_{S}$ and $U(1)_T$ of eq.~(12),
the corresponding K\"{a}hler potential can be chosen to be
invariant.
Then the superpotential should
be invariant up to a constant phase and the gauge
kinetic functions up to imaginary constants.
The coefficient functions $\tilde{K}$ and $Z_{ij}=Z_i\delta_{ij}$
of  the K\"{a}hler potential of eq.~(13)
are largely dominated by PQ-conserving perturbative contributions.
Note that nonperturbative PQ-violating corrections are suppressed
by $q\equiv q_{S,T}$ which are presumed to be less
than $10^{-2}-10^{-3}$.
The same is true for
the gauge kinetic functions $f_a$.
Then up to small corrections of $O(q)$ $\cite{burgess}$,
\bea
\tilde{K}&=&\tilde{K}(S+S^{\ast},T+T^{\ast}),
\nonumber \\
Z_i&=&Z_i(S+S^{\ast},T+T^{\ast}),
\nonumber \\
f_a&=&k_aS+l_aT,
\eea
where $k_a$ and $l_a$ are some real constants.
It is also quite likely that the Yukawa couplings
$\tilde{\lambda}_{ijk}$  which
are relevant for the SUSY phase
$\phi_A={\rm arg}(A_{ijk}/\lambda_{ijk})$ (see eq.~(17))
are also dominated by perturbative contributions,
implying
\beq
\tilde{\lambda}_{ijk}=(h_{ijk}+O(q))\exp (a_{ijk}S+b_{ijk}T),
\eeq
where $h_{ijk}$ is a complex constant while
the constants $a_{ijk}$ and $b_{ijk}$ are real.

We already noted that
$\tilde{W}$
is induced by
nonperturbative  hidden sector dynamics which would be
described  by the
hidden gauge kinetic functions
and the hidden Yukawa couplings
whose forms are restricted by
$U(1)_S$ and $U(1)_T$ as those of eqs. (18) and (19).
Then the arguments based on
$R$-symmetries and dimensional analysis $\cite{amati}$ imply
that $\tilde{W}$ can be written as
\begin{equation}
\tilde{W}=\sum_{n=1}^{N_W} W_n=\sum (1+O(q))d_n e^{(k_{n}S+l_{n}T)},
\end{equation}
where $k_{n}$ and $l_n$ are some {\it real} constants
and $d_{n}$ is a complex constant.

Since the corrections suppressed by a factor
less than $10^{-2}-10^{-3}$
are essentially ignored in our approximation, $\tilde{W}$
includes only the terms such that $|W_n/W_1|\geq 10^{-2}-10^{-3}$
where $W_1$ denotes the term with the largest vacuum value.
The number of such terms, viz $N_W$,  would depend on
some details of  hidden sector dynamics.

Let us briefly discuss $N_W$ for several simple cases.
If the hidden sector contains a simple gauge group ${\cal G}_1$
whose  dynamical mass scale $\Lambda_1$ is far above those of
other groups, then  $\tilde{W}\simeq W_1\sim \Lambda_1^3$
where $W_1$ contains the  gaugino condensation
together with possible matter condensations.
In the case that there exists  another simple
group ${\cal G}_2$ with $\Lambda_2$ comparable to $\Lambda_1$,
$\tilde{W}$ contains  at least two terms
$W_{1,2}\sim \Lambda_{1,2}^3$.
If the gaugino condensations
are largely dominate over other possible contributions,
one simply has $N_W=2$
associated with the two gaugino condensations of ${\cal G}_1$ and ${\cal G}_2$.
Even in the case that matter condensations become important,
if the fields that transform nontrivially
under ${\cal G}_1$ communicate weakly with those
of  ${\cal G}_2$, e.g.
communicate only via nonrenormalizable interactions,
one still has $N_W=2$ but now
$W_1$ and $W_2$ contain both the gaugino and matter condensations
of the ${\cal G}_1$-sector and the ${\cal G}_2$-sector respectively.
As is well known, the simplest case of $N_W=1$ suffers from the runaway
of the dilaton. Because of this,
SUSY breaking in string theories has been discussed
mainly in the context of nonperturbative superpotentials
with $N_W=2$ $\cite{louis1,casas}$.

\section{Small SUSY phases in string-inspired supergravity}

We are now ready to discuss the  mechanism for suppressing
SUSY phases in string-inspired supergravity models.
The coefficients of soft terms of eq.~(17)
show that there are  a number of potentially complex quantities
which can contribute
to the phases
\bea
\phi_A&=&\{{\rm arg}(A_{ijk}/\lambda_{ijk})\},
\nonumber \\
\phi_B&=&\{{\rm arg}(B_{ij}/\mu_{ij})\},\nonumber \\
\phi_C&=&\{{\rm arg}(m_a)\}.
\nonumber
\eea
First of all, the SUSY breaking order parameters $F_I$
can be complex in general.
Furthermore, although $\tilde{K}$ and $Z_i$ are real functions,
their derivatives $\partial_I{\tilde{K}}$,
$\partial_I\bar{\partial}_J\tilde{K}$, and $\partial_IZ_i$
can be complex.  Besides these, we can have
complex $\partial_If_a$, $\partial_I\ln(\tilde{\lambda}_{ijk})$,
and several others.
It is then  convenient
to  classify all the relevant (potentially)
complex quantities as follows:
\bea
X_1 &:&
 \partial_I\tilde{K}, \partial_I\bar{\partial}_J\tilde{K}, \partial_I Z_i,
 \partial_I f_a,
\partial_I\ln(\tilde{\lambda}_{ijk});  \nonumber \\
X_2&:& \partial_I\ln (\tilde{W});
\nonumber \\
X_3&:&\partial_I\ln (\tilde{\mu}_{ij});
\nonumber \\
X_4&:& \partial_I\ln (Y_{ij}), \bar{\partial}_I\ln (Y_{ij}),
\partial_I\bar{\partial}_J\ln (Y_{ij}).
\eea
If  $X_1$ and $X_2$ are all real,
then $\phi_A$ and $\phi_C$ would vanish.
The phase $\phi_B$ is affected also by $X_3$ and $X_4$,
and thus making it small requires more conditions.

Let us consider $X_1$ first.
It is rather easy to see that
all quantities in $X_1$
are {\it real} up to corrections of $O(q)$
 for $\tilde{K}$, $Z_i$,
$f_a$, and $\tilde{\lambda}_{ijk}$ restricted
by  $U(1)_{S,T}$ as eqs. (18) and (19).

Let us consider $X_2=
\partial_I{\rm ln}(\tilde{W})$.
For $\tilde{W}=\sum_1^{N_W}W_n$, one has
$$
X_2=(\sum W_n\partial_I\ln(W_n))/(\sum W_n).
$$
For $W_n$ of eq.~(20),
$\partial_I\ln(W_n)$ is real up to
corrections of $O(q)$.
Then $X_2$ would be real up to  corrections of $O(q)$
if the relative phases ${\rm arg}(W_n/W_m)$   are CP-conserving.

Interestingly enough, for a relatively small value of $N_W$
which would be the most interesting case in view of
its simplicity,
CP-conserving relative phases can be achieved {\it dynamically}.
To see this, let us consider the case
that $\tilde{W}=W_1+W_2$.
In this case of $N_W=2$,
the standard scalar potential in supergravity
gives the following potential
of  the model-independent
axion ${\rm Im}(S)$:
\begin{equation}
V_{\rm axion}=\Omega \,  [\cos ({\rm arg}({W}_2/{W}_1))+O(q)],
\end{equation}
where ${\arg}({W}_2/{W}_1)=(k_2-k_1){\rm Im}(S)+\delta$, and
$\Omega$ and $\delta$
are  real functions which are {\it independent} of ${\rm Im}(S)$.
(See eq.~(20) for the notations.)
Clearly minimizing this axion potential leads to
a CP-conserving value of ${\rm arg}(W_2/W_1)$ up to $O(q)$,
and thus a real $X_2$.
Note that  here $\delta={\rm arg}(d_2/d_1)+(l_2-l_1){\rm Im}(T)$
is of order unity in general, but
it is dynamically relaxed to a CP conserving value
by the vacuum value of ${\rm Im}(S)$. This is quite similar to the
Peccei-Quinn mechanism $\cite{pq}$
in the axion solution to the strong CP problem,
in which $\bar{\theta}$ is dynamically relaxed to zero
by the axion vacuum value.

In the above, we have shown that
$X_1$ and $X_2$ are real up to
corrections of $O(q)$, and thus $\phi_{A,C}=O(q)$,
if  SUSY is broken by the auxiliary components of $S$ and
$T$ which are induced by a nonperturbative
superpotential with $N_W= 2$.
The discussion of the remained phase $\phi_B$
is more model-dependent since
there are a variety of ways $\cite{mu}$
to generate the bilinear terms
$\mu_{ij}\Phi_i\Phi_j$ in the effective superpotential
of eq.~(1).
In the minimal supersymmetric standard model,
the only allowed bilinear term is the so-called
$\mu$-term, $\mu H_1 H_2$, of the two Higgs superfields
$H_1$ and $H_2$.
In the previous section, we
have noted three sources for $\mu$, viz $\mu_{1,2,3}$.
(See eqs. (15) and (16). Here and in the following,
we assume that $\mu H_1H_2$ ($Bh_1 h_2$)
is the only bilinear term
in $W_{\rm eff}$ (${\cal L}_{\rm soft}$), and thus omit $ij$ indices of
the coefficients $\mu_{ij}$ and $B_{ij}$.)
If all of $\mu_{1,2,3}$ give important contributions
to $\mu$, there would not be any reason for $\phi_B$ small.
Thus here  we consider three simple scenarios in which one of
$\mu_{1,2,3}$  dominates over the other two
by a factor larger than  $10^2$ to $10^{3}$.

In the first case that
$\mu_1=\lambda_N\langle N\rangle$ dominates,
$\phi_B$ simply corresponds to $\phi_A$ and thus is of $O(q)$.

In the second case that $\mu_2=(m_{3/2}-\bar{F}_I\bar{\partial}_I)Y$ dominates,
$\phi_B$ would receive additional contribution from $X_4$.
Orbifold compactifications give $Y=0$ $\cite{louis}$ and thus  they
do not correspond to this case.
For Calabi-Yao compactifications, it has been pointed out that
in $(2,2)$ models
$Y$ is related to some Yukawa couplings $\cite{louis}$
(by world sheet Ward identities) which are constants
up to corrections of $O(q)$. This leads to
$X_4=O(q)$ $\cite{ibanez2}$ and thus
$\phi_B=O(q)$.  Thus in the case of $\mu_2$-domination,
 $\phi_B$ can be small
at least in $(2,2)$ Calabi-Yao compactification models

In the third case that $\mu_3$ dominates, $\phi_B$ would receive
a contribution from $X_3=\partial_I\ln (\tilde{\mu})$.
Since $\tilde{\mu}$ is  due to nonperturbative hidden
sector dynamics, using the same arguments applied
for the nonperturbative
superpotential $\tilde{W}$,
it can be written as
$$\tilde{\mu}=\sum \tilde{\mu}_n$$
where  $\tilde{\mu}_n=(1+O(q))z_n\exp (x_n S+y_n T)$.
Here $x_n$ and $y_n$ are some real constants
while $z_n$ is a complex constant.
For hidden sector yielding $\tilde{W}$ with $N_W=2$,
which is the most interesting case for us,
it is expected that $\tilde{\mu}$ also has two terms,
$\tilde{\mu}=
\tilde{\mu}_{1}+\tilde{\mu}_2$.
Again  $\partial_I\ln(\tilde{\mu}_{1,2})$ are real up to $O(q)$.
However to have a real $X_3=\partial_I\ln(\tilde{\mu})$,
one needs the relative phase ${\rm arg}(\tilde{\mu}_1/
\tilde{\mu}_2)$ to be CP conserving.
For $\tilde{W}=W_1+W_2$,
the relative phase  ${\rm arg}(W_1/W_2)$
could be relaxed to a CP conserving value by the
vacuum value of ${\rm Im}(S)$.
For $\tilde{\mu}$, we do not have any such mechanism,
implying $\phi_B$ would {\it not} be small enough unless
further assumptions are made.

The above discussion of $\phi_B$
implies that  perhaps
the first case of $\mu_1$-domination is most attractive
for the purpose of small $\phi_B$,
while the third case of $\mu_3$-domination is
least attractive.

So far, our discussion has been restricted
to the case that SUSY is broken by the
auxiliary components of $S$ and $T$ induced by
a nonperturbative superpotential $\tilde{W}$ with $N_W=2$.
In fact, the discussion can be easily generalized
to more general cases with additional moduli and/or
a larger $N_W$.

Let us suppose an additional modulus $M$
and define the corresponding PQ symmetry
$U(1)_M: M\rightarrow M+i\alpha_M$.
This additional modulus can affect our
previous analysis by two ways:
(a) it can directly
affect SUSY phases
by having a nonzero
auxiliary component $F_M$,
(b) it can affect
$F_I$ ($I=S,T$) via the wave function mixing with $S$ and $T$.
Let $q_M$ denote a factor representing the size of $U(1)_M$-breaking.
Then including  $M$ in the analysis,
it is easy to see that
SUSY phases receive additional contributions
which are  of the order of
either $q_MF_M/m_{3/2}$ or
$q_M\bar{\partial}_I\partial_M\tilde{K}/\bar{\partial
}_I\partial_I\tilde{K}$.

What would be the typical size of the PQ symmetry
breaking factor $q_M$?
For a K\"{a}hler class modulus $M_K$ that is associated
with
the deformation of the K\"{a}hler class of the internal space,
the pseudoscalar component ${\rm Im}(M_K)$ comes from
the zero modes of the antisymmetric tensor field.
Then the corresponding PQ symmetry is
broken  by world sheet instantons $\cite{witten}$,
leading to
$q_{M_K}=e^{-2\pi M_K}$ which can be
small enough, say less than $10^{-2}-10^{-3}$,
for a moderately large value of ${\rm Re}(M_K)$.
For another type of moduli, the complex structure
moduli $M_C$ that is associated with the deformation
of the complex structure, the size of $q_{M_C}$ is somewhat
model-dependent.
For orbifold compactifications, $q_{M_C}$ is still exponentially
small due to the modular symmetry $SL(2,Z)$ $\cite{ibanez1}$.
However for Calabi-Yao cases, $q_{M_C}$ can be  of order
unity even at leading order approximation $\cite{witten}$.
As a result, to achieve small SUSY phases in Calabi-Yao compactification,
one needs to assume that the complex structure moduli
give negligible contribution to SUSY breaking,
viz $F_{M_C}/m_{3/2}\leq 10^{-2}-10^{-3}$, and also have
small wave function mixing with the K\"{a}hler moduli $M_K$,
viz $\bar{\partial}_{M_C}\partial_{M_K} \tilde{K}
/\bar{\partial}_{M_K}\partial_
{M_K}\tilde{K}\leq
10^{-2}-10^{-3}$.

Finally let us consider the dynamical relaxation
of the relative phases  ${\rm arg}(W_n/W_n)$  for $N_W>2$.
Suppose we have $N_A$ axion-like fields $\vec{A}=
(A_1, ..., A_{N_A})$, being the pseudoscalar
components of $S$, $T$ and other possible moduli.
If the PQ symmetries, $U(1)_{\vec{A}}$: $\vec{A}\ra
\vec{A}+\vec{\alpha}$, are good approximate symmetries
as $U(1)_{S,T}$ are, then
the axion-dependence of the
nonperturbative superpotential $\tilde{W}=\sum_1^{N_W}W_n$
would be given by
$W_n\sim e^{i \vec{c}_n\cdot\vec{A}}$
where $\vec{c}_n$ is a real constant vector.
Let $N$ denote the number
of linearly independent vectors among $\{\vec{c}_n-\vec{c}_m\}$.
Obviously  $1\leq N\leq N_A$.
Then analyzing the potential of $\vec{A}$ as we did
in the case of $N_W=2$,  one can see that
if
\beq
N_W\leq N+1,
\eeq
all the relative phases ${\rm arg}(W_n/W_m)$
are relaxed to CP conserving values by the vacuum
values of $\vec{A}$.

\section{Conclusion}

In summary, we have discussed a mechanism for suppressing SUSY
phases in string-inspired supergravity models
in which supersymmetry  is broken
by the  auxiliary components of the dilaton and/or moduli
superfields.
The key ingredients of the
mechanism are: (i) the approximate
PQ symmetries nonlinearly realized for the dilaton and moduli
superfields, (ii) dynamical relaxation of the relative phases
${\rm arg}(W_n/W_m)$ for the nonperturbative
superpotential $\tilde{W}=\sum_1^{N_W} W_n$ of the dilaton
and moduli superfields.

If supersymmetry breaking is dominated
by the dilaton and/or some of the K\"{a}hler class moduli
whose nonperturbative superpotential has $N_W=2$,
which would be perhaps  the
most interesting case in view of its simplicity
(in fact, many of the SUSY breaking scenarios
which have been  discussed
in the context of string theory belong to this category
$\cite{louis1,casas}$),
the SUSY phases $\phi_{A}$ and $\phi_C$
are exponentially suppressed for  moderately
large values of the moduli.
A similar suppression can
occur for the remained phase $\phi_B$ depending upon
how the $\mu$-term is generated.
One would then have SUSY phases {\it less} than $10^{-2}-10^{-3}$
in a quite natural manner.
An interesting feature of this mechanism is that
it is completely independent of how  CP is broken
$\cite{barr}$.
As we have noted, our mechanism for suppressing SUSY phases
can be implemented in  more
general cases with $N_W>2$ if $N_W\leq N_A+1$
for $N_A$ denoting the number of available axion-like fields.


\acknowledgments
This work is supported in part by KOSEF through CTP at Seoul
National University.

\end{document}